\documentclass[a4paper, superscriptaddress, amsfonts, amssymb, amsmath, reprint, showkeys, nofootinbib, twoside,aps,prl]{revtex4-1}
\usepackage[english]{babel}
\usepackage[utf8]{inputenc}
\usepackage{xcolor}
\usepackage{bm}
\usepackage{graphicx}
\usepackage{float}
\usepackage{amsmath}
\usepackage{amsfonts,amssymb,amsthm}
\usepackage[colorinlistoftodos, color=green!40, prependcaption]{todonotes}
\usepackage{amsthm}
\usepackage{mathtools}
\usepackage{physics}
\usepackage{xcolor}
\usepackage{graphicx}
\usepackage[left=23mm,right=13mm,top=35mm,columnsep=15pt]{geometry} 
\usepackage{adjustbox}
\usepackage{placeins}
\usepackage[T1]{fontenc}
\usepackage{lipsum}
\usepackage{amsmath,amssymb,amsthm}

\usepackage{csquotes}

\usepackage{amsmath,amssymb,amsthm}

\usepackage[pdftex, pdftitle={Article}, pdfauthor={Author}]{hyperref} 
\usepackage{multirow}
\usepackage{geometry}
\usepackage{hyperref}
\begin{document}
\title{From Statistical to Structural Synergy: A Predictability Framework to Quantify the Effects due to High-Order Mechanisms}

\author{Yuri Antonacci}
    \email[Correspondence email address: ]{yuri.antonacci@unipa.it}
    \affiliation{Department of Engineering, University of Palermo, Palermo, Italy}
    
\author{Chiara Barà}
    \affiliation{Department of Engineering, University of Palermo, Palermo, Italy}
    \affiliation{Institute of Intelligent Industrial Technologies and Systems for Advanced Manufacturing, National Research Council, Milan, Italy}
    
\author{Laura Sparacino}
    \affiliation{Department of Engineering, University of Palermo, Palermo, Italy}    
    
\author{Daniele Marinazzo}
    \affiliation{Department of Data Analysis, University of Ghent, Belgium}
    
\author{Luca Faes\textsuperscript{\dag}}
    \affiliation{Department of Engineering, University of Palermo, Palermo, Italy}
    \affiliation{Faculty of Technical Sciences, University of Novi Sad, Novi Sad, Serbia}
    
\author{Sebastiano Stramaglia\textsuperscript{\dag}}
    \affiliation{University of Bari Aldo Moro, Bari, Italy}
    \affiliation{Istituto Nazionale di Fisica Nucleare, Sezione di Bari, Bari, Italy}

\begin{abstract}
High-order interactions (HOIs) are increasingly recognized as a hallmark of collective dynamics in complex systems. However, the relationship between high-order behaviours (HOBs), observed as synergistic or redundant statistical dependencies, and high-order mechanisms (HOMs), related to the structural or dynamical rules of the data-generating process, remains difficult to establish from data. Here, focusing on third-order interactions involving two source variables and one target variable, we introduce a predictability-based framework to disentangle these two levels of description in complex network systems. Structural synergy is defined as the excess predictive power gained when two sources are considered jointly beyond the best additive description, and is estimated through polynomial regression by comparing a model with interaction terms against an additive model. Analytical examples and stochastic autoregressive simulations show that statistical dependencies among sources reflecting HOBs can arise even in the absence of HOMs, while synergy due to a non-additive mechanism may remain hidden when the observed synergy-redundancy balance is dominated by redundancy and become detectable only through structural synergy. Applications to climate and source-reconstructed cortical EEG dynamics reveal significant non-additive predictive components despite predominantly redundancy-dominated HOBs. Overall, these findings emphasize that HOBs and HOMs can dissociate: a system may display redundancy-dominated HOBs while still containing a significant synergistic mechanism. The proposed framework therefore supports a more mechanistically informed interpretation of complex-system dynamics and may help to identify when mechanism-based models are needed to predict the response of a system to perturbations or interventions, while also recognizing that, in empirical data, structural synergy should be interpreted as evidence of non-additive predictive structure rather than as a direct identification of the underlying generative mechanism.
\end{abstract}


\maketitle
\footnotetext{\textsuperscript{\dag}These authors contributed equally to this work.}

\section{Introduction}
Complex systems are characterized by collective patterns that arise from the richness of structural and functional interactions among their units. Although graph-based representations can, in principle, accommodate multivariate and non-additive interaction functions defined on node neighbourhoods \cite{peixoto2026graphs}, standard network analyses often rely on pairwise measures of coupling or causality, thereby emphasizing bivariate dependencies while leaving collective effects that cannot be fully characterized through isolated pairwise relations only indirectly described \cite{boccaletti2023structure,fang2024social}.\\
\indent In the field of information theory, high-order effects are commonly described by considering how multiple variables, referred to as sources, jointly contribute to the description or prediction of another variable, referred to as the target \cite{rosas2019quantifying,stramaglia2021quantifying}. Within this framework, high-order interdependencies are tightly connected to the concepts of redundancy and synergy, where the former refers to information about the target that is shared by multiple sources and can therefore be recovered from more than one source separately, while the latter refers to information that becomes available only when sources are considered jointly and cannot be obtained from any source taken in isolation \cite{williams2010nonnegative,stramaglia2024disentangling}.\\
\indent A broad family of approaches has been developed to quantify HOIs from observed multivariate data, and these methods have been applied across several domains, including network physiology \cite{faes2025predictive, sparacino2025decomposing}, cardiovascular and cardiorespiratory dynamics \cite{bara2026embc, mijatovic2024assessing}, neuroscience \cite{antonacci2021measuring,antonacci2024spectral,varley2023multivariate}, and climate networks \cite{faes2025partial}. Overall, these approaches provide principled tools to detect collective effects that remain undetected by standard pairwise analysis, although, since they typically operate at the level of observed statistical dependencies, they primarily characterize high-order behaviours rather than the mechanisms that generate them.\\
\indent This limitation motivates the recent distinction between high-order mechanisms (HOMs) and high-order behaviours (HOBs) \cite{rosas2022disentangling}. HOMs refer to the structural or dynamical rules of the data-generating process, such as Hamiltonians, coupling functions, or dynamical laws containing explicit beyond-pairwise interaction terms \cite{malizia2024reconstructing}. HOBs, instead, refer to emergent statistical patterns in the observed activity that cannot be reduced to the contribution of individual variables or lower-order subsets \cite{faes2022new,rosas2019quantifying}. HOMs and HOBs therefore provide complementary perspectives on high-order phenomena, although their relationship is far from trivial, since simple mechanisms may generate complex HOBs, while complex mechanisms may give rise to different observable high-order effects depending on the system dynamics and operating regime \cite{marinazzo2025behaviors}. Consequently, identifying HOBs can help to constrain the possible underlying mechanisms, but does not guarantee their full characterization.\\
\indent This non-trivial relation becomes particularly evident in Gaussian systems, where statistical synergy can be observed even in the absence of non-linear or non-additive interaction mechanisms. In this setting, mutual-information-based measures can display positive net synergy for linearly related variables, and even when the sources are uncorrelated \cite{barrett2015exploration,bara2026embc}. Recent studies have further clarified this point by showing that genuine group mechanisms can generate distinctive synergistic HOBs, while synergy-dominated HOBs may emerge also from purely pairwise mechanisms under specific structural conditions \cite{robiglio2025synergistic,caprioglio2026synergistic}. These works provide important steps toward linking HOMs and HOBs, but they also highlight the remaining gap: when only observed variables are available, how can one distinguish synergy arising from statistical dependencies among sources from synergy that requires a genuine non-additive source mechanism?\\
\indent In this work, we address this issue within a predictability framework \cite{faes2016predictability,porta2017quantifying,konig2024disentangling} where the behaviour of a network system is modelled in terms of random variables. Specifically, we consider one target and two source variables and first characterize HOBs through a whole-minus-sum (WMS) interaction predictability, which quantifies whether the joint use of the sources improves target prediction beyond the sum of their individual predictive contributions \cite{bara2026embc}. This provides an observed synergy-redundancy balance, but does not by itself establish whether such a collective effect requires a non-additive source mechanism. To address this point, we compare the predictive power obtained from the joint observation of the sources with the predictive power achieved under an additive constraint, where the two sources contribute separately to the target. We refer to the difference between these two predictive powers as \textit{structural synergy}, since it identifies the component of target predictability that cannot be reproduced by additive source effects and can therefore be associated with a non-additive source mechanism, corresponding to HOMs as intended in the present work.\\
\indent The proposed framework is validated across progressively more realistic settings. We first derive analytical results in linear Gaussian and non-linear systems, and then test the method in stochastic autoregressive systems with controlled additive and non-additive mechanisms and source dependencies. We further demonstrate the broad applicability of the framework in two real-world scenarios: climate system dynamics described by interactions among representative indices of El Niño and the El Niño–Southern Oscillation (ENSO), the dominant mode of interannual variability in sea surface temperature and atmospheric pressure over the equatorial Pacific Ocean; and brain networks probed through source reconstructed electroencephalographic (EEG) signals recorded during motor execution. Together, these analyses show how the proposed framework can support the interpretation of emergent high-order effects in complex network systems by separating dependency-driven behaviours from non-linear, non-additive interaction mechanisms.

\section{Methods}

 \subsection{High-order behaviours and mechanisms}
\label{sec:defHOMHOB}

Let us consider a network system 
$\mathcal{S}=\{Y,X_1,\ldots,X_{N_s}\}$, 
where $Y$ is a target random variable and 
$\boldsymbol{X}=[X_1,\ldots,X_{N_s}]$ is the vector of source variables. In this setting, the dependence of the target on the sources can be generally described as
\begin{equation}
    Y = \phi(\boldsymbol{X}) + U,
\end{equation}
where $\phi$ is an unknown, possibly non-linear function describing how the sources contribute to the target, and $U \perp \boldsymbol{X}$ accounts for the component of $Y$ not explained by the sources.

HOMs refer to structural or dynamical rules through which the sources influence the target via genuine group interactions that cannot be reduced to independent source effects or lower-order relations \cite{rosas2022disentangling,robiglio2025synergistic}. Accordingly, the absence of HOMs with respect to the target Y corresponds to the case in which the effect of the sources on the target can be decomposed into separate source-specific contributions that combine additively:
\begin{equation}
    Y = h_1(X_1)+\cdots+h_{N_s}(X_{N_s}) + U,
\end{equation}
where each $h_i(X_i)$ represents a possibly non-linear contribution of the individual source $X_i$ to the target. When this additive representation is not sufficient, for instance because cross-terms or other non-additive interactions between sources are required, the system is said to possess HOMs with respect to $Y$.

HOBs, instead, refer to statistical or predictive patterns in the observed variables that cannot be reduced to the contribution of individual variables or lower-order subsets \cite{mcgill1954multivariate,rosas2022disentangling}. They describe the collective dependence of the target on the sources, without specifying whether this dependence originates from a genuine non-additive mechanism or from statistical dependencies among the sources; for this reason, the detection of HOBs does not, by itself, imply the presence of HOMs. In the following, we focus on three-node systems with two sources ($N_s=2$), denoted as $X_1$ and $X_2$, and one target $Y$.

\subsection{Predictability measures of HOBs and HOMs}
\label{sec:predictability_HOM_HOB}

Let $X_1$ and $X_2$ be two source random variables and $Y$ the target random variable, jointly distributed according to an unknown probability distribution. We quantify the predictive contribution of the sources through the mean squared prediction error, considering predictions based on the two sources jointly, on each source separately, and on an additive restriction of the joint prediction.

For any measurable predictor
$f: X_1 \times X_2 \rightarrow \mathbb{R}$,
we define the quadratic risk functional
\begin{equation}
    R[f] =
    \mathbb{E}\!\left[(Y-f(X_1,X_2))^2\right],
\end{equation}
which represents the mean squared prediction error (MSPE) obtained when predicting $Y$ from the joint observation of $(X_1,X_2)$. Similarly, for predictors depending on each source separately, we define
\begin{align}
    R_1[g_1] &=
    \mathbb{E}\!\left[(Y-g_1(X_1))^2\right],\\
    R_2[g_2] &=
    \mathbb{E}\!\left[(Y-g_2(X_2))^2\right],
\end{align}
where $g_1:X_1\rightarrow\mathbb{R}$ and $g_2:X_2\rightarrow\mathbb{R}$ are measurable predictors of $Y$ based on the individual source variables $X_1$ and $X_2$, respectively. The optimal joint predictor, minimizing the MSPE over all measurable functions of $(X_1,X_2)$, is the conditional expectation
\begin{equation}
    f^*(X_1,X_2)=\mathbb{E}[Y\mid X_1,X_2],
\end{equation}
with minimum MSPE \cite{hastie2009elements}
\begin{equation}
    R^* =
    R[f^*] =
    \mathbb{E}\!\left[(Y-\mathbb{E}[Y\mid X_1,X_2])^2\right].
\end{equation}
Restricting prediction to each source separately yields
\begin{align}
    g_1^*(X_1) &= \mathbb{E}[Y\mid X_1],\\
    g_2^*(X_2) &= \mathbb{E}[Y\mid X_2],
\end{align}
with minimum MSPEs $R_1^*=R_1[g_1^*]$ and $R_2^*=R_2[g_2^*]$.

Let $\varepsilon=Y-\mathbb{E}[Y\mid X_1,X_2]$ be the residual error of the optimal joint predictor. By the law of total variance \cite{blitzstein2019introduction}, and exploiting the fact that $\mathbb{E}[\varepsilon\mid X_1,X_2]=0$, we can write
\begin{equation}
    \sigma_Y^2 = \lambda + R^*,
\end{equation}
where $R^*=\sigma_\varepsilon^2$ is the unexplained variance and
\begin{equation}
    \lambda = \sigma_Y^2 - R^*
\end{equation}
is the predictive power explained by the joint knowledge of $X_1$ and $X_2$. Analogously, the individual predictive powers associated with the two sources are
\begin{equation}
    \lambda_1 = \sigma_Y^2-R_1^*,
    \qquad
    \lambda_2 = \sigma_Y^2-R_2^* .
\end{equation}

Following the WMS rationale \cite{timme2014synergy} in the framework of predictability \cite{porta2017quantifying,faes2025predictive}, we compute the interaction predictability as
\begin{equation}
    \Delta
    =
    \lambda-\lambda_1-\lambda_2
    =
    R_1^*+R_2^*-R^*-\sigma_Y^2 .
    \label{DELTA}
\end{equation}
This quantity describes the observed synergy-redundancy balance at the level of HOBs: $\Delta>0$ indicates that the joint observation of the sources provides a predictive contribution larger than the sum of the individual contributions, whereas $\Delta<0$ indicates a redundancy-dominated balance.

To isolate the component of predictability that cannot be reproduced by separate source effects, we introduce the additive predictive class
\begin{equation}
    \mathcal{H}_{\mathrm{add}}
    =
    \left\{
    h(X_1,X_2)=h_1(X_1)+h_2(X_2)
    \right\}.
\end{equation}
The optimal additive predictor is
\begin{equation}
    h^*
    =
    \arg\min_{h\in\mathcal{H}_{\mathrm{add}}} R[h],
\end{equation}
with minimum MSPE
\begin{equation}
    R_A^* = R[h^*].
\end{equation}
The predictive power explained by the best additive model is
\begin{equation}
    \lambda_A = \sigma_Y^2-R_A^*,
\end{equation}
and the corresponding additive interaction predictability is
\begin{equation}
    \Delta_A
    =
    \lambda_A-\lambda_1-\lambda_2
    =
    R_1^*+R_2^*-R_A^*-\sigma_Y^2 .
    \label{DELTA-A}
\end{equation}

Since the additive class is a restriction of the class of all measurable predictors, $R^*\leq R_A^*$ and therefore $\Delta\geq\Delta_A$. Structural synergy is then defined as
\begin{equation}
    S_s
    =
    \Delta-\Delta_A
    =
    \lambda-\lambda_A=R_A^*-R^* .
    \label{SS}
\end{equation}
Thus, while $\Delta$ describes the observed HOB, $S_s$ quantifies the part of target predictability that is lost when source effects are constrained to be additive. Equivalently, a positive value of $S_s$ indicates that the predictive contribution of the sources cannot be fully reproduced by separate source-specific effects, but requires a non-additive combination of the sources. In controlled generative systems, this non-additive predictive component can be directly related to the presence of a non-additive source-target mechanism. In observational applications, however, $S_s$ should be interpreted more cautiously as evidence that the target contains a non-additive predictive component with respect to the selected sources and model class, rather than as direct evidence identifying the underlying mechanism.

The regimes obtained by comparing $\Delta$ and $\Delta_A$ are summarized in Table~\ref{tab:presence}. When $S_s=0$, the observed HOB is fully reproduced by the additive benchmark, whereas $S_s>0$ reveals a non-additive contribution to target predictability. Importantly, this structural component may be present even when $\Delta<0$, corresponding to a masked structural synergy regime in which the system appears redundancy-dominated at the HOB level while still exhibiting a non-additive predictive component. In controlled settings, this regime can be interpreted as evidence of a non-additive mechanism that remains hidden behind redundancy-dominated HOBs.

\begin{table*}[t]
\caption{Classification of interaction regimes based on the comparison between the interaction predictability $\Delta$ and its additive counterpart $\Delta_A$.}
\centering
\small
\begin{tabular}{l |c| c| c| c| c}
   \hline
   \hline
   Regime & $\Delta$ & $\Delta_A$ & $S_s$ & HOB & HOM \\
   \hline
   No high-order effect
   & $0$ & $0$ & $0$ & Absent & Absent \\

   Dependency-driven synergy
   & $>0$ & $=\Delta$ & $0$ & Synergistic & Absent \\
   
   Dependency-driven redundancy
   & $<0$ & $=\Delta$ & $0$ & Redundant & Absent \\

   Visible structural synergy
   & $>0$ & $<\Delta$ & $>0$ & Synergistic & Present \\

   Masked structural synergy
   & $<0$ & $<\Delta$ & $>0$ & Redundant & Present \\
   \hline
   \hline
\end{tabular}
\label{tab:presence}
\end{table*}

\subsection{Practical evaluation of structural synergy} 
\label{practical}

Let $Y$ be a target random variable and $X_1$ and $X_2$ two source random variables, all assumed to be zero-mean. To approximate the optimal predictors introduced above, we consider linear regression problems on polynomial regressors, formulated here in terms of random variables \cite{brillinger2011generalized}. In particular, four regression models are considered, corresponding to the residual variances required to compute $\Delta$, $\Delta_A$, and, consequently, $S_s$ according to Eqs.~(\ref{DELTA})-(\ref{SS}). The MATLAB codes and functions implementing the proposed framework are publicly available at \href{https://github.com/YuriAntonacci/HOB2HOM}{https://github.com/YuriAntonacci/HOB2HOM}.

First, the target $Y$ is predicted using only a single source, either $X_1$ or $X_2$:
\begin{equation}
Y = \sum_{k=1}^{p} \alpha_{i,k} X_{i}^k + U_{i}, \qquad i\in \{1,2\},
\end{equation}
where $X_i^k = (X_i)^k$ denotes the $k^{\textrm{th}}$ power of the $i^{\textrm{th}}$ source variable. The coefficients $\{\alpha_{i,k}\}$ weight the contribution of each polynomial term $X_i^k$ in the prediction of $Y$ based on $X_i$ up to order $p$, while $U_i$ is the corresponding residual term with variance $R_i^*=\mathbb{E}[U_i^2]$.

Then, a full bivariate model including polynomial interaction terms is considered:
\begin{equation} 
\label{full_model}
Y =
\sum_{k=1}^{p} \beta_k X_{1}^k
+ \sum_{k=1}^{p} \gamma_k X_{2}^k
+ \sum_{k=1}^{p-1} \sum_{j=1}^{p-k} \delta_{k j} X_{1}^k X_{2}^j
+ U,
\end{equation}
where $U$ is the corresponding residual with variance $R^*=\mathbb{E}[U^2]$, and the double summation includes all interaction terms with total degree $j+k \le p$. Lastly, the additive bivariate model is defined by excluding interaction terms:
\begin{equation}
Y = \sum_{k=1}^{p} \tilde{\beta}_k X_{1}^k
+ \sum_{k=1}^{p} \tilde{\gamma}_k X_{2}^k
+ U_A,
\end{equation}
where $U_A$ is the corresponding residual term with variance $R_A^*=\mathbb{E}[U_A^2]$.

In all cases, the integer $p$ denotes the maximum order of the polynomial expansion and controls the dimensionality of the projection space. The regression coefficients are estimated from observed data via least-squares regression \cite{lutkepohl2013introduction}, yielding sample-based approximations of the residual variances required to compute $\Delta$ and $\Delta_A$ according to Eqs.~(\ref{DELTA}) and (\ref{DELTA-A}), and structural synergy $S_s$ according to Eq.~(\ref{SS}).

\subsection{Statistical assessment of structural synergy}

To assess the statistical significance of structural synergy, a surrogate data analysis approach is employed \cite{theiler1992testing,palus1997detecting}. The surrogate procedure is used for two related but distinct purposes: first, to test whether the estimated value of $S_s$ at a fixed polynomial order is larger than expected under the null hypothesis of no non-additive predictive contribution; second, to assess whether increasing the polynomial order from $p-1$ to $p$ provides a statistically significant improvement in the description of the interaction structure.

For the first test, surrogate realizations are generated by randomly permuting the samples of the polynomial interaction terms appearing in Eq.~\ref{full_model}, while leaving the individual source terms unchanged. This procedure destroys the correspondence between the interaction terms, the additive regression terms, and the target variable, thereby testing the null hypothesis that the interaction terms do not provide an additional non-additive contribution to target prediction, i.e., $S_s=0$. The procedure is repeated $N_{\mathrm{surr}}$ times, and $S_s$ is computed for each surrogate realization, yielding a surrogate distribution under the null hypothesis. The significance threshold is then defined as the \(100(1-\alpha)\)-th percentile of this distribution, where $\alpha$ denotes the chosen significance level. The original value of $S_s$ is considered statistically significant if it exceeds this threshold.

\indent For the second test, a greedy search is performed to identify the polynomial order that most adequately describes the non-additive source interactions contributing to target prediction. The estimation strategy described above is applied for increasing values of $p$, and the search is stopped when moving from order $p-1$ to order $p$ does not yield a statistically significant improvement. To test the contribution of the newly introduced order-$p$ terms, the samples of the polynomial terms present at order $p$ but absent from the model of order $p-1$ are randomly shuffled, while all lower-order terms are kept unchanged. These newly introduced terms include the additive monomials $X_1^p$ and $X_2^p$, as well as the interaction monomials $X_1^kX_2^j$ with total degree $k+j=p$. This procedure tests the null hypothesis that the order-$p$ polynomial terms do not provide an additional contribution to structural synergy beyond that already captured by the lower-order model. Repeating the procedure $N_{\mathrm{surr}}$ times yields a surrogate distribution for the order-$p$ model, from which the significance threshold is defined as the $100(1-\alpha)$-th percentile. If the value of $S_s$ computed on the original data at order $p$ exceeds this threshold, the newly introduced polynomial terms are considered to provide a statistically significant improvement, and the model of order $p$ is retained as a more adequate description of the interaction structure.

\section{Validation on theoretical examples}
In this section, the proposed approach for detecting structural synergy is illustrated through two theoretical examples designed to exhibit distinct configurations of source interactions. These examples highlight three key theoretical properties of the proposed decomposition: (i) interaction predictability may be positive even in the absence of genuine interaction mechanisms, arising solely from statistical dependencies between the sources; (ii) structural synergy isolates the contribution of non-additive source interactions by removing dependency-driven effects; and (iii) source correlation plays a crucial role in shaping the balance between redundancy and synergy, leading to qualitatively distinct interaction regimes. In both theoretical examples, all quantities entering the definition of $S_s$ are derived analytically from the statistical structure of the underlying random variables. \\
\indent We consider additive-noise models of the form $Y = f(X_1, X_2) + U$, where $U$ is a zero-mean Gaussian noise term uncorrelated with $(X_1, X_2)$. The sources $(X_1, X_2)$ are assumed jointly Gaussian with zero mean, unit variance ($\mathbb{E}[X_1]=\mathbb{E}[X_2]=0$, $\sigma^2_{X_1}=\sigma^2_{X_2}=1$), and correlation $\mathbb{E}[X_1X_2]=r_{12}$, with $r_{12}\in(-1,1)$. Two paradigmatic cases are analyzed: a linear additive model $f(X_1,X_2)=aX_1+bX_2$ and a non-additive generative model $f(X_1,X_2)=cX_1X_2$. The former represents a setting where interaction predictability may arise exclusively from statistical dependencies between the sources, implying vanishing structural synergy. The latter embodies a genuine mechanistic interaction, for which structural synergy is strictly positive. In both cases, the source correlation $r_{12}$ critically modulates the balance between redundancy and synergy, highlighting the interplay between dependency-driven and mechanism-driven effects.\\
\indent Before analyzing these examples, we note that statistical consistency requires the relevant covariance matrices to be positive definite \cite{konig2024disentangling}. In the linear Gaussian model $Y=aX_1+bX_2+U$, the vector $(Y,X_1,X_2)$ is jointly Gaussian and its covariance matrix is given by
\[
\Sigma_{[YX_1X_2]}= \begin{pmatrix}
1 & a+b\,r_{12} & ar_{12}+b\\
a+b\,r_{12} & 1 & r_{12}\\
ar_{12}+b & r_{12} & 1
\end{pmatrix},
\]
with $|\Sigma_{[YX_1X_2]}|=(1-r_{12}^2)\big[1-(a^2+b^2+2ab\,r_{12})\big]$. Requiring this matrix to be positive definite imposes constraints on the parameters $(a,b,r_{12})$, in particular $|\Sigma_{[YX_1X_2]}|>0$ requires $|r_{12}|<1$ and $a^2+b^2+2ab\,r_{12}<1$.

In the model $Y=cX_1X_2+U$, no Gaussianity assumption can be imposed on the target variable $Y$, whose distribution is induced by the nonlinear transformation of the sources. In this case, the only covariance matrix that must be positive definite by construction is that of the jointly Gaussian sources defined as
\[
\Sigma_{[X_1X_2]}=\begin{pmatrix}
1 & r_{12}\\
r_{12} & 1
\end{pmatrix},
\]
which requires $|r_{12}|<1$ since $|\Sigma_{[X_1X_2]}|=1-r_{12}^2$. Throughout the
following analysis, we therefore restrict attention to source correlations satisfying $|r_{12}|<1$.

\subsection{Linear Gaussian system} 
\begin{figure*}
    \centering    \includegraphics{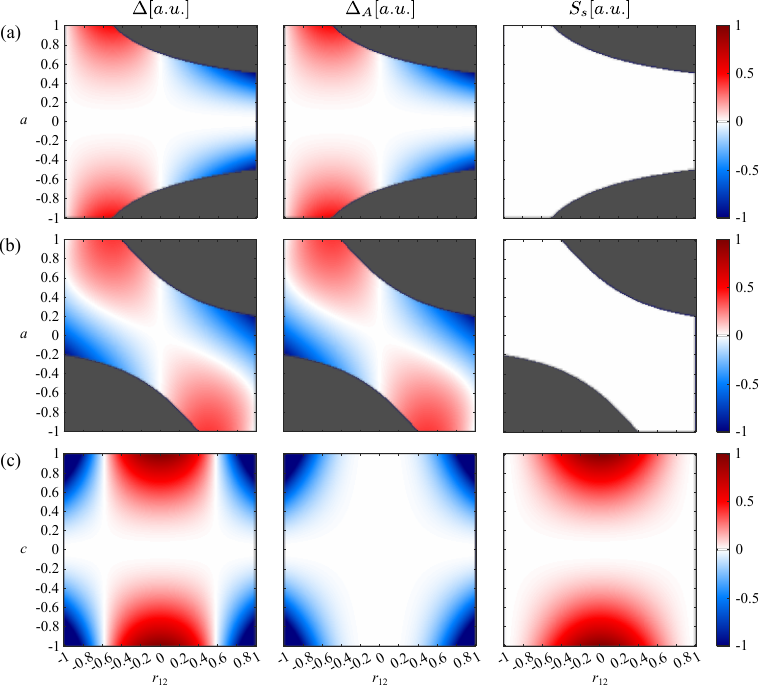}
    \caption{Theoretical trends of interaction predictability ($\Delta$ and $\Delta_A$) and structural synergy ($S_s$) for a linear Gaussian system (a,b) and a non-additive multiplicative system (c), shown as a function of the source correlation $r_{12}$. Different settings of the model parameters are considered: (a) linear Gaussian case with $a=b$, with $a,b\in(-1,1)$; (b) linear Gaussian case with $a\in(-1,1)$ and $b=0.8$; and (c) non-additive multiplicative system, with $c\in(-1,1)$ and varying $r_{12}$.}
    \label{fig:gaussian}
\end{figure*}

Since jointly Gaussian systems are fully characterized by linear conditional expectations \cite{barrett2015exploration}, $\Delta$ and $\Delta_A$ can be derived directly from the optimal linear predictors based on the individual and joint sources. In particular, for the linear Gaussian model $Y=aX_1+bX_2+U$ the optimal predictors are defined as $f^*(X_1,X_2)=\mathbb{E}[Y\mid X_1,X_2]=aX_1+bX_2$, where $R^*=\sigma_U^2$. Restricting prediction to the sources $X_1$ and $X_2$ yields $g_1^*=\mathbb{E}[Y\mid X_1]=aX_1+b\mathbb{E}[X_2|X_1]=(a+br_{12})X_1$, since for Gaussian variables $\mathbb{E}[X_2|X_1]=r_{12}X_1$ \cite{barrett2015exploration}. The corresponding MSPE is $R^*_1=b^2(1-r_{12}^2)+\sigma_U^2$. Analogously, the MSPE associated with prediction from $X_2$ is $R^*_2=a^2(1-r_{12}^2)+\sigma_U^2$. Given that $\mathbb{E}[Y]=0$, the relevant variance is defined as $\sigma_Y^2=\mathbb{E}[Y^2]=\mathbb{E}[(aX_1+bX_2+U)^2]=a^2+b^2+2abr_{12}+\sigma_U^2$ by just exploiting the fact that $\mathbb{E}[X_1U]=\mathbb{E}[X_2U]=0$. Substituting these quantities into Eq.~(\ref{DELTA}) yields:
\begin{equation}
    \Delta=-r_{12}^2(a^2+b^2)-2abr_{12}. \label{GAUSS}
\end{equation} 
The interaction predictability $\Delta$ is a quadratic function of the source correlation $r_{12}$, and the equation $\Delta(r_{12})=0$ admits two roots, $r_{12}=0$ and $r_{12}=-2ab/(a^2+b^2)$. 
Accordingly, $\Delta$ is positive only in the intermediate regime where $r_{12}$ lies strictly between the two roots, corresponding to the red regions in Fig.~\ref{fig:gaussian}(a,b). The extent of this region depends on the coefficients $a$ and $b$, as illustrated by the two parameter settings varied in panels (a) and (b), and occurs when the source correlation counteracts the linear coefficients, i.e., when $ab,r_{12}<0$ and $|r_{12}|<|2ab|/(a^2+b^2)$. Since the optimal predictor $\mathbb{E}[Y\mid X_1,X_2]$ is linear and additive in the sources, restricting the prediction to the additive function class does not increase the MSPE, so that $R_A^*=R^*$, $\Delta_A=\Delta$, and $S_s=\Delta-\Delta_A=0$. Therefore, positive values of $\Delta$ in this linear Gaussian setting correspond to the \textit{dependency-driven synergy} regime of Table~\ref{tab:presence}: the observed HOB is synergy-dominated, but no HOM is present because the effect is entirely induced by source dependencies. Conversely, outside this interval $\Delta$ becomes negative, yielding a \textit{dependency-driven redundancy} regime in which the HOB is redundancy-dominated while structural synergy remains absent. At the roots, where $\Delta=\Delta_A=S_s=0$, the system falls into the \textit{no high-order effect} condition reported in Table~\ref{tab:presence}.

\subsection{Non-linear system with multiplicative source interaction}
We now consider a model defined by $Y = cX_1X_2 + U$ where the optimal predictor is given by $f^*(X_1,X_2)=\mathbb{E}[Y\mid X_1,X_2]=cX_1X_2$, while the corresponding minimum MSPE is $R^*=\sigma_U^2$. Restricting the prediction to the single sources yields
$\mathbb{E}[Y\mid X_1]=cr_{12}X_1^2$ and $\mathbb{E}[Y\mid X_2]=cr_{12}X_2^2$,
leading to $R_1^*=R_2^*=c^2(1-r_{12}^2)+\sigma_U^2$ and $\sigma_Y^2=c^2(1+r_{12}^2)+\sigma_U^2$ (derivations in Appendix). Substituting these quantities into
Eq.~(\ref{DELTA}) yields
\begin{equation}
    \Delta=c^2(1-3r_{12}^2). \label{delta_nonlin}
\end{equation}
Since $c^2 \ge 0$, the sign of $\Delta$ is entirely determined by the factor $1-3r_{12}^2$. 
Specifically, $\Delta>0$ for $|r_{12}|<1/\sqrt{3}$, $\Delta=0$ for $|r_{12}|=1/\sqrt{3}$ (or trivially for $c=0$), and $\Delta<0$ for $|r_{12}|>1/\sqrt{3}$. 
This behavior is illustrated in Fig.~\ref{fig:gaussian}(c), where the positive values of $\Delta$ occur in the central region of the $r_{12}$ domain, while negative values emerge as the source correlation increases in magnitude.

To assess whether such interaction predictability can be explained without
invoking genuine interaction mechanisms, we compute the best additive predictor
$h(X_1,X_2)=d+h_1(X_1)+h_2(X_2)$ and the corresponding additive interaction
predictability $\Delta_A$ (full derivation in Appendix).
This yields the closed-form expression
\begin{equation}
    \Delta_A=R_1^*+R_2^*-R_A^*-\sigma_Y^2=c^2(1-3r_{12}^2)-c^2\frac{(1-r_{12}^2)^2}{1+r_{12}^2},
\end{equation}
and hence the structural synergy
\begin{equation}
    S_s=\Delta-\Delta_A=c^2 \frac{(1-r_{12}^2)^2}{1+r_{12}^2}. \label{ss_nonlin}
\end{equation}

Overall, the nonlinear system exhibits the regimes involving structural synergy summarized in Table~\ref{tab:presence}, as illustrated in Fig.~\ref{fig:gaussian}(c). When $c\neq 0$ and $|r_{12}|<1/\sqrt{3}$, the system operates in a \textit{visible structural synergy} regime, characterized by a synergy-dominated HOB ($\Delta>0$) together with a strictly positive structural synergy ($S_s>0$), implying that $\Delta_A<\Delta$. As the source correlation increases beyond the transition point, i.e., for $1/\sqrt{3}< |r_{12}|<1$, the system enters a \textit{masked structural synergy} regime: interaction predictability becomes negative ($\Delta<0$), indicating a redundancy-dominated HOB, while structural synergy remains strictly positive ($S_s>0$), showing that the non-additive mechanism is still present but masked by source dependence. The boundary case $|r_{12}|=1/\sqrt{3}$ corresponds to $\Delta=0$ with $S_s>0$, and therefore marks the transition between the visible and masked structural synergy regimes. Finally, when $c=0$, the model reduces to noise-driven dynamics, for which the optimal joint predictor is additive and no high-order effect is present, yielding $\Delta=\Delta_A=S_s=0$, consistently with the first row of Table~\ref{tab:presence}.

\section{Validation in a stochastic autoregressive system}
After establishing the analytical behaviour of the proposed measures, we next evaluate their practical estimation in a stochastic autoregressive system to provide a closer bridge between the analytical examples and the real-world applications, which are based on time series that may exhibit temporal autocorrelations \cite{bara2026embc}. In this controlled dynamical context, additive source effects, non-additive source interactions, and source dependence can be independently controlled, allowing us to test whether the polynomial-regression implementation of the framework recovers the expected distinction between dependency-driven HOBs and genuine non-additive mechanisms.

A system composed of two autoregressive processes $X_1$ and $X_2$, interacting to determine the dynamics of a third process $Y$, is considered. Specifically, the system is defined by the following equations \cite{kilian2017structural, terasvirta2010modelling}:
\begin{subequations}
    \begin{alignat}{3}
        X_{1,n} &= 0.8 X_{1,n-1}+0.2e_{1,n},\\
        X_{2,n} &= 0.6 X_{2,n-1}+ r_{12} X_{1,n-1}+0.2e_{2,n},\\
        Y_{n} &= 0.5 Y_{n-1}+cX_{1,n-1}X_{2,n-1} \\
        &+c_AX_{1,n-1}+c_AX_{2,n-1}+0.2e_{3,n}, \nonumber \label{Y_NAR}
    \end{alignat}
\end{subequations}
where $e_i$, $i=1,\dots,3$, are uncorrelated Gaussian noises with zero mean and unit variance, $c$ regulates the strength of the non-additive interaction between $X_1$ and $X_2$ in determining $Y$, $c_A$ controls the additive dependence of $Y$ on the two sources, and $r_{12}$ regulates the dependence of $X_2$ on $X_1$.

Fifty realizations of the three processes, each with length 1000 samples, were generated under four simulation settings designed to separately probe the effects of non-additive coupling, additive coupling, and source dependence. In the first two settings, the sources were kept independent by fixing $r_{12}=0$, while the non-additive coupling strength $c$ was varied in the range $[0:0.2:2]$. The first setting isolated the non-additive contribution by setting $c_A=0$, whereas the second setting allowed additive and non-additive effects to coexist by varying them complementarily, with $c_A=2-c$, thereby progressively changing their relative contribution to the target dynamics. In the remaining two settings, the coupling strengths were fixed and the source-dependence parameter $r_{12}$ was varied in the range $[-1:0.2:1]$: the third setting tested a purely non-additive mechanism under increasing source dependence by setting $c=2$ and $c_A=0$, while the fourth setting considered the complementary purely additive case, obtained by setting $c=0$ and $c_A=2$.

$\Delta$, $\Delta_A$, and $S_s$ were computed by taking the current state of the process $Y$, i.e., $Y_n$, as the target variable, and the lagged states of the processes $X_1$ and $X_2$, i.e., $X_{1,n-1}$ and $X_{2,n-1}$, as source variables. In all cases, the polynomial order of the predictive models was fixed to $p=2$, consistently with the quadratic interaction term defining the system. The statistical significance of $S_s$ was assessed by generating 100 surrogate realizations and applying the percentile-based procedure described in the Methods section, with significance level $\alpha=0.05$.

\begin{figure}
    \centering
    \includegraphics[width=\linewidth]{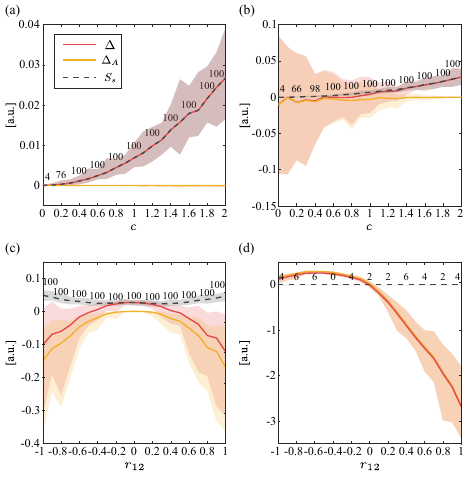}
    \caption{Estimated trends of interaction predictability ($\Delta$, red lines), additive interaction predictability ($\Delta_A$, yellow lines), and structural synergy ($S_s$, dashed black lines) for the stochastic autoregressive system. Lines report averages across realizations, while shaded areas indicate the corresponding minimum--maximum range. Panels (a) and (b) refer to uncorrelated sources ($r_{12}=0$) with $c\in[0,2]$, considering $c_A=0$ and $c_A=2-c$, respectively. Panels (c) and (d) refer to source-dependent systems with $r_{12}\in[-1,1]$, considering a non-additive target mechanism ($c=2$, $c_A=0$) and a purely additive target mechanism ($c=0$, $c_A=2$), respectively. The numbers reported in the panels indicate the percentage of realizations for which $S_s$ was deemed statistically significant according to surrogate data analysis.}
    \label{fig:AR}
\end{figure}

Fig.~\ref{fig:AR} reports the trends of $\Delta$, $\Delta_A$, and $S_s$ across the four investigated configurations of the stochastic autoregressive system, shown as mean values across the fifty generated realizations together with the corresponding minimum--maximum range. Overall, the results reproduce the interaction regimes summarized in Table~\ref{tab:presence}, showing that HOBs and HOMs may either coexist or dissociate depending on the source dependence and on the structure of the generative model.

In panels (a) and (b), the sources are uncorrelated ($r_{12}=0$) and the non-additive coupling strength $c$ is progressively increased. In this condition, $\Delta$ and $S_s$ increase with $c$, while $\Delta_A$ remains, on average, close to zero, consistently with a visible structural synergy regime in which the observed HOB is synergy-dominated ($\Delta>0$) and supported by a genuine non-additive mechanism ($S_s>0$). In panel (b), where additive and non-additive effects coexist, larger values of $c_A$ increase the variability of $\Delta_A$ across realizations, but do not alter the overall interpretation, since significant values of $S_s$ remain associated with the presence of the non-additive interaction term.

Panels (c) and (d) assess the effect of source dependence by varying $r_{12}$. In panel (c), where the target is driven by the non-additive interaction between the sources, $S_s$ remains statistically significant across different values of $r_{12}$, indicating the persistence of the underlying mechanism. However, as $|r_{12}|$ increases, $\Delta$ decreases and becomes non-positive, while $\Delta_A$ follows the same trend with lower values than $\Delta$. This corresponds to the transition from visible to masked structural synergy: the non-additive mechanism is still detected through $S_s>0$, but the observed HOB is no longer synergy-dominated because source dependence induces redundancy. Conversely, in panel (d), where the target depends only on additive source effects, $S_s$ remains close to zero and is not statistically significant for all values of $r_{12}$, while the positive and equal values of $\Delta$ and $\Delta_A$ observed for negatively correlated sources reflect dependency-driven synergy, i.e., a synergy-dominated HOB that does not require a non-additive mechanism. \\
\indent Overall, these results support the practical reliability of the proposed framework, since the estimated trends of $\Delta$, $\Delta_A$, and $S_s$ reproduce the expected regimes across all configurations and show that the method can distinguish dependency-driven HOBs from genuine non-additive mechanisms. In particular, $S_s$ consistently detects structural synergy when the target is generated by a non-additive interaction between the sources, even when source dependence modifies the sign and magnitude of the HOBs quantified by $\Delta$. This distinction is especially relevant when $\Delta$ and $\Delta_A$ are strongly affected by source dependence or exhibit large variability across realizations, since relying only on interaction predictability could lead to ambiguous conclusions about the presence of synergy or redundancy \cite{bara2026embc}. By contrast, the structural synergy term remains statistically significant only when a non-additive interaction mechanism is present, supporting its use as a more specific marker of mechanism-driven synergy.

\section{Application to real-world scenarios}

To assess the applicability of the proposed framework beyond controlled simulations, we consider two real-world scenarios drawn from markedly different domains: climate dynamics and source reconstructed brain activity during motor execution. These systems are characterized by multivariate interactions, strong dependencies among observed variables, and emergent behaviours that are not necessarily reducible to pairwise relations. Moreover, both scenarios have already been investigated in the literature using information-theoretic approaches, which revealed the presence of HOBs in terms of synergistic and redundant interdependencies among system units. Specifically, in the climate application, we analyze interactions among large-scale climatic indices involved in the El Niño--Southern Oscillation, a system previously shown to exhibit redundant and synergistic dynamic effects among climatological variables \cite{faes2025partial,stramaglia2024disentangling}. In the EEG application, we focus on source reconstructed motor-network dynamics, where spectral information-theoretic analyses have highlighted hierarchically organized interactions across individual, pairwise, and higher-order levels during motor execution \cite{antonacci2021measuring,antonacci2024spectral}. These two scenarios therefore provide suitable empirical test cases for the present framework: rather than merely detecting the presence of HOBs, we ask whether the observed synergy-redundancy patterns contain a component that cannot be reproduced by additive source effects alone.

\subsection{Climate case study}

We investigate a representative case study in climate science concerning the most important interannual climate variability on Earth, i.e., the periodic fluctuation in sea surface temperature and atmospheric air pressure in correspondence with the equatorial Pacific Ocean (El Niño and the Southern Oscillation, ENSO) \cite{mcphaden2006enso}. ENSO warm and cold phases are characterized by two main indicators: sea surface temperature anomalies in the east-central tropical Pacific (NINO34) and standardized surface air pressure difference between Tahiti and Darwin, known as the Southern Oscillation Index (SOI). These variables, however, are embedded within a broader network of interacting climate variables including e.g. the Tropical Southern Atlantic Index (TSA), the Pacific Decadal Oscillation (PDO), and the North Tropical Atlantic (NTA) \cite{stramaglia2024disentangling}.

In this framework, we explore the potential presence of HOBs and HOMs by analyzing different triplets of these climatological indices. Specifically, SOI was treated as the target variable $Y$, while different pairs of the other indices were used as sources $\{X_1,X_2\}$, i.e., \{NINO34, TSA\}, \{NINO34, PDO\}, \{NINO34, NTA\}, \{TSA, PDO\}, \{TSA, NTA\}, and \{PDO, NTA\}. The data are obtained from a publicly available dataset \cite{silini2023assessing}, which provides monthly climatological indices over the period 1950–2016, resulting in a total of 792 observations. Specifically, the Spearman's rank correlation coefficient was evaluated for each pair of sources and then $\Delta$, $\Delta_A$, and $S_s$ were computed with $p \in \{2,3,4\}$. The statistical significance of the computed $S_s$ values, as well as of the value of $p$ employed for its practical computation, was assessed by generating one-hundred surrogate datasets ($N_{\mathrm{surr}}=100$) and using the percentile-based approach ($\alpha=0.05$) to find the significance threshold as described in the methods section. 

\begin{figure}
    \centering
    \includegraphics[width=\linewidth]{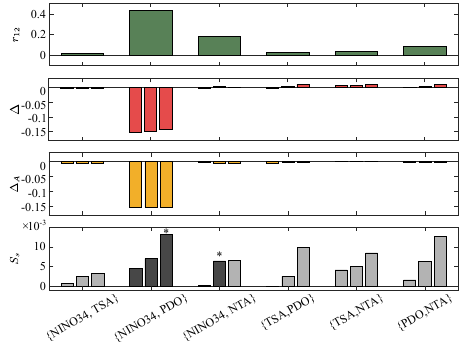}
    \caption{Bar plots reporting the Spearman correlation coefficient $r_{12}$ between the two sources (top row), and the values of interaction predictability $\Delta$, additive interaction predictability $\Delta_A$, and structural synergy $S_s$ (bottom row), computed using SOI as the target and the source pairs indicated on the x-axis. The measures were estimated for polynomial orders $p\in\{2,3,4\}$, represented by grouped bars. Dark gray bars indicate values of $S_s$ that were statistically significant according to surrogate analysis. Asterisks denote a statistically significant increase in $S_s$ when increasing the polynomial order from $p-1$ to $p$; these markers are reported only when $S_s$ itself is statistically significant.}
    \label{fig:clima}
\end{figure}

Fig.~\ref{fig:clima} reports the values of $\Delta$, $\Delta_A$, and $S_s$ obtained for the six configurations defined from the considered climatic indices, using SOI as target and all possible pairs among NINO34, TSA, PDO, and NTA as sources. Overall, the results suggest that HOIs are not uniformly distributed across the analysed climate network. For most triplets, both $\Delta$ and $\Delta_A$ remain close to zero and no statistically significant values of $S_s$ are observed, suggesting the absence of clear HOBs and HOMs in these configurations. 

A different pattern emerges for the source pairs including NINO34, in particular $\{\mathrm{NINO34},\mathrm{PDO}\}$ and $\{\mathrm{NINO34},\mathrm{NTA}\}$. In these cases, the HOBs are redundancy-dominated or weakly expressed, as indicated by negative or near-zero values of $\Delta$. Nevertheless, $S_s$ reaches statistically significant values, indicating that the joint predictability of SOI contains a non-additive component that cannot be reproduced by additive source effects alone. According to the classification in Table~\ref{tab:presence}, these configurations are therefore consistent with a masked structural synergy regime: source dependencies induce a redundancy-dominated HOB, while significant structural synergy indicates the presence of a non-additive predictive component compatible with an underlying HOM. This behaviour is in agreement with the theoretical examples and stochastic simulations reported above, where source dependence can reduce the value of $\Delta$ and mask the synergy-dominated signature of a non-additive mechanism.

The interpretation of these findings is supported by previous analyses performed on the same set of climatic indices. Previous studies have suggested a central role of ENSO-related variables, identifying NINO34 and SOI as pivotal nodes in climate networks reconstructed through causality-based measures, with relevant connections also involving PDO and NTA \cite{silini2023assessing}. Moreover, focusing on the information flow from NINO34 to SOI, it has recently been shown that, although most of this transfer can be attributed to a dyadic contribution, a non-negligible fraction is due to many-body effects \cite{stramaglia2024disentangling}. Finally, the application of partial information decomposition for random processes to the same ENSO-related dataset showed that source pairs including NINO34 share the largest amount of dynamic information with SOI, with NTA playing a relevant role in both redundant and synergistic dynamic components \cite{faes2025partial}.

Taken together, these results suggest that the structural synergy detected here is concentrated in configurations that previous information-theoretic studies had already identified as relevant for ENSO dynamics. Our analysis adds a complementary perspective: while previous approaches revealed the presence of statistical dependencies or HOBs among ENSO-related variables, the present framework asks whether the observed predictability patterns contain components that cannot be reproduced by additive source effects alone. In this sense, the significant $S_s$ observed for the NINO34-PDO and NINO34-NTA source pairs suggests that part of the predictability of SOI requires non-additive combinations of the selected sources. This finding is compatible with the presence of non-additive interactions among ENSO-related climatic indices, although the empirical nature of the data prevents the identification of the underlying physical processes from $S_s$ alone. Conversely, the absence of significant effects in the remaining triplets indicates that non-additive predictive components are not widespread across all combinations of indices, but are selectively associated with specific ENSO-related source configurations.

\subsection{Application to source reconstructed EEG dynamics during motor execution}

The dataset comprises EEG signals recorded from 100 healthy participants using 64 electrodes referenced to both mastoids according to the 10–10 system, with a sampling frequency of $f_s = 160$ Hz \cite{schalk2004bci2000,PhysioNet}. The data are publicly available at \href{https://physionet.org/content/eegmmidb/1.0.0/}{https://physionet.org/content/eegmmidb/1.0.0/}. The experimental protocol included ten trials of two-minute runs of a motor execution task. During the TASK condition, a visual target was presented on both sides of the screen, and participants were instructed to cyclically open and close both fists until the target disappeared; this was followed by a relaxation period (REST condition).

EEG preprocessing was performed using the EEGLAB toolbox \cite{delorme2004eeglab} in MATLAB (The MathWorks 2025b). Raw signals were detrended and band-pass filtered using a second-order Butterworth filter (1–45 Hz). Then, brain current source density distributions were extracted for each participant using the eLORETA software, which exploits a discrete linear weighted minimum-norm inverse solution \cite{pascual2007discrete}. Brain source activity was reconstructed within the cortical gray matter (6239 isotropic voxels with 5 mm spatial resolution) in the MNI152 space \cite{mazziotta2001probabilistic}. Five regions of interest (ROIs) within the motor network were selected based on previous literature \cite{antonacci2024spectral,pirovano2023rehabilitation}, including the primary motor cortex (M$_1$) and premotor cortex (pMC) in both the left (L) and right (R) hemispheres, as well as the supplementary motor area (SMA). All voxels within an 8 mm radius from each seed were assigned to the corresponding ROI, with the constraint of non-overlapping regions; in cases of overlap, voxels were assigned to the nearest centroid region. The eLORETA inverse solution was computed using a regularization parameter $\lambda = 0.05$ \cite{pirovano2023rehabilitation}. Five source reconstructed time series were obtained by averaging the magnitude of source activity across all voxels within each ROI. Source power values were log-transformed to improve normality and reduce skewness. Finally, the ROI signals were epoched into 10 temporal windows per condition, trial, and participant, each lasting 4 s, corresponding to 640 samples, a window length chosen to favour local stationarity while preserving enough samples for the subsequent predictability-based analysis \cite{antonacci2024spectral,antonacci2021measuring}.

Three ROI triplets within the motor network were investigated considering as source variables $X_1=\{LM_1\}$, and $X_2=\{RM_1\}$ and as target $Y \in \{SMA, LpMC, RpMC\}$. This choice was motivated by the established involvement of bilateral primary motor cortices, supplementary motor area, and premotor cortices in motor execution \cite{grefkes2008dynamic}. By keeping the two primary motor regions as common sources and varying the target within higher-level motor areas, this configuration allowed us to assess whether non-additive source interactions are preferentially expressed toward specific nodes of the motor network. In particular, the triplet targeting the SMA was expected to capture a core component of motor-network integration, whereas the premotor targets were included to test whether similar effects also extend to lateral premotor regions, which are involved in motor planning and execution but may play a less central integrative role \cite{grefkes2008dynamic}. 

Each time series was first z-scored. Then, for each subject, trial, and experimental condition, $\Delta$, $\Delta_A$, and $S_s$ were computed using polynomial orders $p \in \{2,3,4\}$, while Spearman's rank correlation coefficient was evaluated between $RM_1$ and $LM_1$. The statistical significance of $S_s$ was assessed by generating 100 surrogate datasets and, to account for multiple testing across the ten trials, the corresponding significance threshold was determined using a Bonferroni-corrected level of $\alpha = 0.05/10$. The statistical significance of the increase in $S_s$ obtained by increasing the model order $p$ was also assessed through surrogate analysis, using a significance level of $\alpha = 0.05$. For each subject and condition, the values of the computed measures were averaged across trials, while statistical significance was assigned at the subject level whenever at least one trial showed a significant effect. Differences between REST and TASK conditions were evaluated using the paired Wilcoxon signed-rank test with $\alpha = 0.05$.

\begin{figure}
    \centering
    \includegraphics[width=\linewidth]{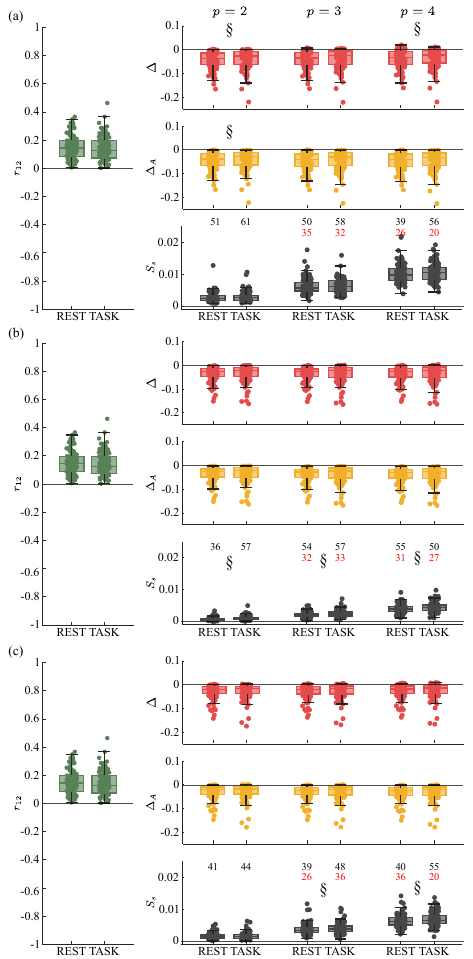}
    \caption{Boxplots and individual subject values, averaged across trials, of the Spearman correlation coefficient $r_{12}$, interaction predictability $\Delta$, additive interaction predictability $\Delta_A$, and structural synergy $S_s$ for three ROI triplets: (a) $Y=\mathrm{SMA}$, $X_1=\mathrm{LM1}$, and $X_2=\mathrm{RM1}$; (b) $Y=\mathrm{LpMC}$, $X_1=\mathrm{LM1}$, and $X_2=\mathrm{RM1}$; and (c) $Y=\mathrm{RpMC}$, $X_1=\mathrm{LM1}$, and $X_2=\mathrm{RM1}$. Results are reported for polynomial orders $p\in\{2,3,4\}$ in both REST and TASK conditions. Black numbers above each distribution indicate the percentage of subjects for whom $S_s$ was statistically significant according to surrogate analysis with multiple-comparison correction. Red numbers indicate the percentage of subjects for whom increasing the polynomial order from $p-1$ to $p$ yielded a statistically significant increase in $S_s$. The symbol $\S$ denotes a statistically significant difference between REST and TASK distributions.}
    \label{fig:brain}
\end{figure}

Fig.~\ref{fig:brain} reports the values of Spearman's correlation coefficient between the two sources, together with $\Delta$, $\Delta_A$, and $S_s$, for the three investigated ROI triplets within the motor network. Across all configurations, the correlation between the two source regions remains positive and weak, with values around $0.1$, and does not show clear modulation between REST and TASK conditions. Thus, the changes observed in the predictability measures cannot be ascribed to a strong task-dependent modulation of source correlation. For all the triplets and polynomial orders, both $\Delta$ and $\Delta_A$ are predominantly negative, denoting the presence of redundant HOBs. This suggests that the two reconstructed source signals provide partially overlapping information about the target. However, this redundancy-dominated regime is accompanied by non-zero values of $S_s$, which are statistically significant in a substantial fraction of subjects, especially during TASK. This pattern indicates that the apparent redundancy observed through $\Delta$ may coexist with a structural synergistic component, suggesting a non-additive predictive component that is not directly apparent from the HOB measure $\Delta$. \\
\indent These findings can be interpreted in light of the theoretical non-linear model and the stochastic autoregressive simulations reported above. In those examples, positive dependence between the sources can drive both $\Delta$ and $\Delta_A$ toward negative values, producing redundancy-dominated HOBs even when a genuine non-additive mechanism is present. A similar situation appears in the EEG analysis: the motor network displays redundancy-dominated HOBs, but this does not exclude the presence of structural synergy. Rather, the significant values of $S_s$ suggest a masked structural synergy regime, in which source dependencies and overlapping predictive contributions may hide the synergistic mechanism at the level of $\Delta$. Task-related effects further support this interpretation since motor task execution is associated with a reduction of redundancy, reflected by less negative values of $\Delta$ and/or $\Delta_A$, together with an increase of $S_s$ as the polynomial order increases. This trend is particularly evident when SMA is considered as target (panel \textit{a}), where the modulation of $S_s$ is not accompanied by a direct modulation of $\Delta$ or $\Delta_A$, suggesting that the task mainly affects the non-additive component rather than the overall synergy-redundancy balance. In the same configuration, the largest absolute values of $S_s$ are observed, although the REST-TASK modulation does not always reach statistical significance. This interpretation is consistent with the central integrative role of the SMA within the motor network, as also suggested by effective-connectivity studies showing that SMA participates in the task-dependent modulation of intra- and interhemispheric motor interactions \cite{grefkes2008dynamic}. It is also in line with previous information-theoretic analyses of the same motor network, where HOIs were found to be predominantly redundancy-dominated and to decrease during motor execution \cite{antonacci2024spectral}. 

The present analysis refines this interpretation by showing that redundancy-dominated HOBs may coexist with a structural synergistic component. In other words, although previous results described the organization of HOIs in the motor network in terms of redundancy and synergy, the proposed framework suggests that part of these high-order effects may be supported by non-additive source mechanisms. This provides a possible explanation for the coexistence of negative values of $\Delta$ and $\Delta_A$ with significant values of $S_s$: the observed motor-network dynamics remain redundancy-dominated in terms of HOBs, while still exhibiting masked structural synergy compatible with a non-additive predictive organization. \\
\indent From a physiological perspective, this interpretation is plausible since the analyzed ROIs belong to the core of the motor network, including bilateral primary motor cortices, premotor cortices, and SMA. Dynamic causal modeling studies have shown that even simple hand movements involve context-dependent intra- and interhemispheric interactions among these regions, with the SMA playing a key role in promoting or suppressing activity in M1 and in coordinating motor-network dynamics \cite{grefkes2008dynamic}. Therefore, the presence of significant $S_s$ in the investigated triplets is compatible with non-linear or non-additive coordination within functionally relevant motor pathways, which may become more evident during task execution despite the persistence of redundancy-dominated HOBs. This interpretation should also take into account the intrinsic limitations of EEG source reconstruction, since, although working at the source level reduces the effect of sensor-space mixing, the propagation of neural activity through the electromagnetic forward model cannot be fully inverted, and residual field spread or source leakage may still affect the estimated dependencies among ROI time series \cite{van2019critical}. Accordingly, the significant values of $S_s$ should be interpreted as evidence of non-additive predictive organization among source-reconstructed cortical signals, rather than as direct proof of anatomically isolated non-additive neural mechanisms.

\section{Conclusions}

This work introduces a predictability-based framework to assess the extent to which HOBs can be attributed to HOMs, rather than to mere statistical dependencies among source variables. Inspired by explainable machine-learning approaches for quantifying feature contributions \cite{konig2024disentangling}, the proposed framework exploits the WMS principle to quantify the synergy--redundancy balance observed at the level of HOBs. This balance is first evaluated without restrictions, through the interaction predictability $\Delta$, and then under an additive constraint, through its additive counterpart $\Delta_A$. Structural synergy is then introduced to measure the excess predictive power gained by allowing non-additive interactions between the sources, thereby providing a direct way to separate dependency-driven HOBs from mechanism-driven synergistic effects. These properties make our approach intrinsically different from most existing approaches proposed to investigate HOIs: information-theoretic measures such as partial information decomposition \cite{williams2010nonnegative}, O-information \cite{rosas2019quantifying}, or related extensions to random processes \cite{faes2022new,faes2025partial} primarily characterize the statistical organization of the observed variables, thus quantifying HOBs in terms of synergistic or redundant interdependencies. Other recent approaches have investigated the relationship between HOMs and HOBs by showing either how known HOMs generate synergistic high-order effects in controlled models, or how synergy-dominated HOBs may arise even from purely pairwise mechanisms under specific structural conditions \cite{robiglio2025synergistic,caprioglio2026synergistic,bara2026embc}. The present framework addresses this relationship from a complementary perspective, providing a predictability-based criterion to distinguish dependency-driven HOBs from mechanism-driven synergistic effects, operationalized in a computationally efficient way using polynomial regression. 

Both the theoretical examples and the stochastic autoregressive simulations showed that positive interaction predictability may arise even in the absence of non-additive mechanisms, while genuine non-additive mechanisms may remain hidden behind redundancy-dominated HOBs when the sources are statistically dependent. We refer to this latter condition as masked structural synergy, which represents a central result of the proposed framework: a system may display $\Delta<0$, and thus appear redundant in terms of HOBs, while still exhibiting significant structural synergy. Importantly, the stochastic simulations confirmed that this distinction can be recovered in finite-sample dynamical data, with $S_s$ remaining sensitive to non-additive mechanisms even when $\Delta$ and $\Delta_A$ are strongly affected by source dependence.

The two explanatory applications further supported this interpretation. In the climate case study, significant structural synergy was selectively observed for ENSO-related source configurations involving NINO34, consistently with previous information-theoretic analyses of the same climatic indices. In the EEG application, source reconstructed dynamics were predominantly characterized by redundancy-dominated HOBs, in agreement with previous studies of HOIs in the motor network. Nevertheless, significant values of $S_s$ revealed the presence of masked structural synergy, suggesting that non-additive source mechanisms may coexist with, and be hidden by, redundant HOBs during motor execution. Overall, these findings emphasize that HOBs and HOMs can dissociate: a system may display redundancy-dominated HOBs while still containing a significant synergistic mechanism. This distinction is relevant not only for interpreting complex-system dynamics, but also for anticipating their response to perturbations. Observed HOBs may be sufficient for prediction in some settings, but interventions and control strategies require understanding which predictive effects are compatible with dependency-driven organization and which instead point to non-additive source mechanisms. Structural synergy therefore provides a bridge between data-driven characterization and mechanism-based modelling, by identifying configurations in which a purely additive description is not sufficient to account for the observed predictability.

Despite the merits of the proposed approach, some limitations should be acknowledged. First, the framework is still based on the WMS principle, which is known to conflate synergy and redundancy and may inflate redundancy due to multiple counting effects, especially as the number of sources increases \cite{mediano2025toward}. A natural extension would therefore be to exploit partial information decomposition (PID), which explicitly separates synergistic from redundant contributions. By unmasking the synergistic component of the observed statistical dependence, PID could detect structural synergy, especially when non-additive predictive effects are partially hidden by redundancy. This gain in interpretability would, however, come at the cost of the arbitrariness inherent to PID decompositions, whose results depend on the adopted redundancy function and on the assumptions underlying the decomposition \cite{williams2010nonnegative}. Second, the practical implementation based on polynomial regression may suffer from a rapid growth in the number of terms when increasing either the polynomial order or the number of sources, making the approach computationally demanding and potentially unstable beyond moderate model orders. A further limitation concerns the interpretation of structural synergy in observational data. While $S_s$ identifies predictive components that cannot be reproduced by additive source effects, it does not by itself identify the underlying mechanism. Therefore, in empirical applications, structural synergy should be regarded as evidence of non-additive predictive organization, whose mechanistic interpretation requires additional modelling assumptions, experimental perturbations, or independent physiological/physical knowledge.\\
\indent Future developments should therefore extend the framework beyond the two-source case, investigate alternative non-linear predictive models, and formalize dynamic versions of structural synergy for random processes. Such extensions may provide a broader bridge between predictability-based approaches, information-theoretic decompositions, and mechanistic modelling of HOIs in complex systems.

\bibliography{reference.bib}

\appendix
\appendix
\setcounter{equation}{0}
\renewcommand{\theequation}{A.\arabic{equation}}
\renewcommand{\theHequation}{A.\arabic{equation}} 

\section{Appendix}\label{app:additive_full}

\section{Derivation of the predictive terms for the non-additive multiplicative system}

\subsection{Individual and joint predictive terms}

When one predictor is considered, it is necessary to compute the term
$\mathbb{E}[Y\mid X_1]$, which can be written as
\begin{equation}
\mathbb{E}[Y\mid X_1]
=\mathbb{E}[cX_1X_2\mid X_1]+\mathbb{E}[U\mid X_1]
=cX_1\mathbb{E}[X_2\mid X_1],
\end{equation}
where $U\perp X_1$ and $\mathbb{E}[f(X_1)\mid X_1]=f(X_1)$. Thus, it is well-known
that for jointly Gaussian variables with zero mean and unit variance \cite{barrett2015exploration}
\begin{equation}
\mathbb{E}[X_2\mid X_1]
=\frac{\mathbb{E}[X_2X_1]}{\sigma_{X_1}^2}\,X_1
=r_{12}X_1,
\end{equation}
which yields $\mathbb{E}[Y\mid X_1]=cr_{12}X_1^2$ (and analogously for $X_2$).

A further step is the computation of the relevant MSPE
$R_1^*=\mathbb{E}[(Y-\mathbb{E}[Y\mid X_1])^2]$. Starting from
\begin{equation}
Y-\mathbb{E}[Y\mid X_1]
=cX_1X_2+U-cr_{12}X_1^2
=cX_1(X_2-r_{12}X_1)+U,
\end{equation}
we define the random variable $W=X_2-r_{12}X_1$, which is Gaussian with
$\mathbb{E}[W]=0$ and variance $\mathbb{E}[W^2]=1-r_{12}^2$. Indeed,
\begin{equation}
\begin{split}
\mathbb{E}\!\left[(X_2-r_{12}X_1)^2\right]
&=\mathbb{E}[X_2^2-2X_2X_1r_{12}+r_{12}^2X_1^2]\\
&=\sigma_{X_2}^2-2r_{12}\mathbb{E}[X_2X_1]+r_{12}^2\sigma_{X_1}^2\\
&=1-r_{12}^2.
\end{split}
\end{equation}
Moreover, $W\perp X_1$ since $\mathbb{E}[X_1W]=\mathbb{E}[X_1X_2-r_{12}X_1^2]=0$.
Therefore,
\begin{equation}
\begin{split}
\mathbb{E}\!\left[(cX_1W+U)^2\right]
&=\mathbb{E}[c^2X_1^2W^2]+\mathbb{E}[2cX_1WU]+\mathbb{E}[U^2]\\
&=c^2\mathbb{E}[X_1^2]\mathbb{E}[W^2]+\sigma_U^2\\
&=c^2\sigma_{X_1}^2(1-r_{12}^2)+\sigma_U^2\\
&=c^2(1-r_{12}^2)+\sigma_U^2,
\end{split}
\end{equation}
so that $R_1^*=c^2(1-r_{12}^2)+\sigma_U^2$ (and symmetrically $R_2^*=R_1^*$).

The variance of $Y$ can be retrieved from the computation of all statistical
moments for $\mathbb{E}[(cX_1X_2+U-cr_{12})^2]$ as follows:
\begin{equation}
\begin{split}
&=\mathbb{E}[c^2X_1^2X_2^2+U^2+c^2r_{12}^2]\\
&\quad + \mathbb{E}[2cX_1X_2U-2c^2X_1X_2r_{12}-2Ucr_{12}]\\
&=c^2\mathbb{E}[X_1^2X_2^2]+\sigma_U^2+c^2r_{12}^2-2c^2r_{12}^2\\
&=c^2(1+r_{12}^2)+\sigma_U^2,
\end{split}
\end{equation}
which follows from Isserlis' theorem \cite{konig2024disentangling}, since
\begin{equation}
\mathbb{E}[X_1^2X_2^2]
=\mathbb{E}[X_1^2]\mathbb{E}[X_2^2]+2(\mathbb{E}[X_1X_2])^2
=1+2r_{12}^2.
\end{equation}
Thus,
\begin{equation}
\sigma_Y^2=c^2(1+r_{12}^2)+\sigma_U^2.
\end{equation}

\subsection{Best additive predictor and additive MSPE}
\label{app:additive_full}

We herein derive the computation of $\Delta_A$ by finding the best additive predictor
$h(X_1,X_2)=d+h_1(X_1)+h_2(X_2)$ (where $d\in\mathbb{R}$ is an intercept and
$\mathbb{E}[h_1(X_1)]=\mathbb{E}[h_2(X_2)]=0$) in the sense of minimizing
\begin{equation}
R_A^*
=\min_{d,h_1,h_2\in L^2}\mathbb{E}[(Y-d-h_1(X_1)-h_2(X_2))^2].
\end{equation}
According with \cite{buja1989linear}, differentiating with respect to $d$ yields
\begin{equation}
\frac{\partial}{\partial d}R_A[h_1,h_2,d]
=-2\mathbb{E}[Y-d-h_1(X_1)-h_2(X_2)],
\end{equation}
which leads to $d^*=\mathbb{E}[Y]=cr_{12}$. With $d=d^*$ fixed, define
$V=Y-d^*-h_1(X_1)-h_2(X_2)$; the minimizer is characterized by the normal
equations \cite{buja1989linear}:
\begin{subequations}
\label{eq:normal_equations}
\begin{align}
\mathbb{E}[V^*\mid X_1] &=0, \label{eq:normal_eq_x1}\\
\mathbb{E}[V^*\mid X_2] &=0, \label{eq:normal_eq_x2}
\end{align}
\end{subequations}
where $V^*:=Y-d^*-h_1^*(X_1)-h_2^*(X_2)$. Expanding \eqref{eq:normal_equations}
gives the functional equations
\begin{subequations}
\label{eq:functional_eqs}
\begin{align}
h_1^*(X_1)
&=\mathbb{E}[Y-d^*\mid X_1]-\mathbb{E}[h_2^*(X_2)\mid X_1],
\label{eq:h1_star_general}\\
h_2^*(X_2)
&=\mathbb{E}[Y-d^*\mid X_2]-\mathbb{E}[h_1^*(X_1)\mid X_2].
\label{eq:h2_star_general}
\end{align}
\end{subequations}
Let us introduce $Z:=X_1X_2-r_{12}$, so that, since $d^*=cr_{12}$, we have $Y-d^*=cZ+U$, with $\mathbb{E}[Z]=0$. Defining $V_{h_1,h_2}=Y-d^*-h_1(X_1)-h_2(X_2)$, and taking into account that $U\perp (X_1,X_2)$ ($\mathbb{E}[U]=0$) the additive risk can be written as
\begin{align}
\label{eq:RA_decomposition}
&\mathbb{E}\!\left[
\left(Y-d^*-h_1(X_1)-h_2(X_2)\right)^2
\right] \nonumber\\
&=
\mathbb{E}\!\left[
\left(cZ-h_1(X_1)-h_2(X_2)\right)^2
\right]
+\mathbb{E}[U^2].
\end{align}
Therefore,
\begin{equation}
\label{eq:RA_star_reduce}
R_A^*
=\sigma_U^2
+\min_{h_1,h_2\in L^2}
\mathbb{E}\!\left[(cZ-h_1(X_1)-h_2(X_2))^2\right].
\end{equation}

Taking into account that $(X_1,X_2)$ is jointly Gaussian with unit variances and
correlation $r_{12}$, the conditional expectations are linear,
\begin{equation}
\mathbb{E}[X_2\mid X_1]=r_{12}X_1,\qquad
\mathbb{E}[X_1\mid X_2]=r_{12}X_2,
\end{equation}
and, in particular,
\begin{equation}
\label{eq:EZ_given_X1}
\mathbb{E}[Z\mid X_1]=r_{12}(X_1^2-1),
\end{equation}
and symmetrically
\begin{equation}
\label{eq:EZ_given_X2}
\mathbb{E}[Z\mid X_2]=r_{12}(X_2^2-1).
\end{equation}
Equations \eqref{eq:EZ_given_X1}--\eqref{eq:EZ_given_X2} 
show that the component of the centered
interaction term $Z$ that is predictable from each individual source is
quadratic in that source. Therefore, the additive projection of $Z$ can
be represented in the span of the centered quadratic functions
$X_1^2-1$ and $X_2^2-1$. This choice is consistent with the
characterization of the best additive $L^2$ approximation, whose
residual must be orthogonal to all functions of each source separately \cite{konig2024disentangling}.
We therefore write
\begin{equation}
    h_1(X_1)=c\alpha(X_1^2-1), \qquad
    h_2(X_2)=c\beta(X_2^2-1). \label{eq:hypothesis_space}
\end{equation}
Let $A:=X_1^2-1$ and $B:=X_2^2-1$. Minimizing \eqref{eq:RA_star_reduce} over
\eqref{eq:hypothesis_space} is equivalent to
\begin{equation}
\label{eq:J_ab}
\min_{\alpha,\beta\in\mathbb{R}} \;
J(\alpha,\beta)
:=
\mathbb{E}\!\left[(Z-\alpha A-\beta B)^2\right].
\end{equation}
The normal equations for \eqref{eq:J_ab} are obtained by setting
$\partial_\alpha J=\partial_\beta J=0$, yielding
\begin{subequations}
\label{eq:normal_alpha_beta}
\begin{align}
\mathbb{E}\!\left[A\big(Z-\alpha A-\beta B\big)\right] &= 0,
\label{eq:norm_ab_1}\\
\mathbb{E}\!\left[B\big(Z-\alpha A-\beta B\big)\right] &= 0.
\label{eq:norm_ab_2}
\end{align}
\end{subequations}

Using Isserlis' theorem for jointly Gaussian variables:
\begin{align}
\mathbb{E}[A^2]
&=\mathbb{E}[(X_1^2-1)^2]
=\mathbb{E}[X_1^4]-2\mathbb{E}[X_1^2]+1
=2,\\
\mathbb{E}[B^2] &= 2,\\
\mathbb{E}[AB]
&=\mathbb{E}[X_1^2X_2^2]-1
=(1+2r_{12}^2)-1
=2r_{12}^2,\\
\mathbb{E}[AZ]
&=\mathbb{E}[(X_1^2-1)(X_1X_2-r_{12})]
=2r_{12},\\
\mathbb{E}[BZ] &= 2r_{12}.
\end{align}

Substituting the moments into \eqref{eq:normal_alpha_beta} yields the linear
system
\begin{equation}
\begin{bmatrix}
2 & 2r_{12}^2\\
2r_{12}^2 & 2
\end{bmatrix}
\begin{bmatrix}
\alpha\\ \beta
\end{bmatrix}
=
\begin{bmatrix}
2r_{12}\\ 2r_{12}
\end{bmatrix}.
\end{equation}
Dividing by $2$ and exploiting symmetry gives $\alpha=\beta$, hence
\begin{equation}
\alpha(1+r_{12}^2)=r_{12}
\quad\Rightarrow\quad
\alpha^*=\beta^*=\frac{r_{12}}{1+r_{12}^2}.
\end{equation}
Therefore, the optimal additive predictors are
\begin{subequations}
\label{eq:h1h2_star_final}
\begin{align}
h_1^*(X_1) &= c\,\frac{r_{12}}{1+r_{12}^2}\,(X_1^2-1),\\
h_2^*(X_2) &= c\,\frac{r_{12}}{1+r_{12}^2}\,(X_2^2-1).
\end{align}
\end{subequations}
Thus, the best additive predictor can be written as
\begin{equation}
h^*(X_1,X_2)
= cr_{12}
+c\,\frac{r_{12}}{1+r_{12}^2}\Big[(X_1^2-1)+(X_2^2-1)\Big].
\end{equation}
The minimum value of \eqref{eq:J_ab} is equal to the variance of $Z$ minus the
amount of variance explained by its optimal additive approximation. Substituting
the Gaussian moments computed above, we obtain
\begin{equation}
J^*
=\mathbb{E}[Z^2]-\frac{4r_{12}^2}{1+r_{12}^2}
=\frac{(1-r_{12}^2)^2}{1+r_{12}^2}.
\end{equation}
Finally, from \eqref{eq:RA_decomposition}--\eqref{eq:RA_star_reduce},
\begin{equation}
R_A^*
=\sigma_U^2+c^2\,\frac{(1-r_{12}^2)^2}{1+r_{12}^2}.
\end{equation}
By recalling Eq.~(\ref{DELTA-A}) we obtain
\begin{equation}
\Delta_A
=R_1^*+R_2^*-R_A^*-\sigma_Y^2
=c^2(1-3r_{12}^2)-c^2\,\frac{(1-r_{12}^2)^2}{1+r_{12}^2},
\end{equation}
and substituting into Eq.~(\ref{SS}) yields
\begin{equation}
S_s=\Delta-\Delta_A
=c^2\,\frac{(1-r_{12}^2)^2}{1+r_{12}^2}.
\end{equation}

\end{document}